# Structural and transport properties of GaAs/δ-Mn/GaAs/In$_x$Ga$_{1-x}$As/GaAs quantum well

B.A. Aronzon[1,6], M.V. Kovalchuk[1], E.M. Pashaev[1], M.A. Chuev[2,1], V.V. Kvardakov[1], I.A. Subbotin[3,1], V.V. Rylkov[1], M.A. Pankov[1,6], A.S. Lagutin[1], B.N. Zvonkov[4], Yu.A. Danilov[4], O.V. Vihrova[4], A.V. Lashkul[5], R. Laiho[5]

[1]*Russian Research Center "Kurchatov institute", Kurchatov sq. 1, 123182, Moscow, Russia*
[2]*Institute of Physics & Technology of RAS, Nakhimovskii Ave. 34, 117218 Moscow, Russia*
[3]*A.V. Shubnikov Institute of Crystallography of RAS, Leninskii Ave. 59, 119333 Moscow, Russia*
[4] *Physico-Technical Research Institute of Nizhny Novgorod University, 603950, N. Novgorod, Russia*
[5]*Wihuri Physical Laboratory, Department of Physics, University of Turku, FIN-20014 Turku, Finland*
[6]*P.N. Lebedev Research Center in Physics, Leninskii Ave. 53, 119991 Moscow, Russia*

**ABSTRACT**

We report results of investigations of structural and transport properties of GaAs/Ga$_{1-x}$In$_x$As/GaAs quantum wells (QWs) having a 0.5-1.8 ML thick Mn layer, separated from the QW by a 3 nm thick spacer. The structure has hole mobility of about 2000 cm$^2$/(V·s) being by several orders of magnitude higher than in known ferromagnetic two-dimensional structures. The analysis of the electro-physical properties of these systems is based on detailed study of their structure by means of high-resolution X-ray diffractometry and glancing-incidence reflection, which allow us to restore the depth profiles of structural characteristics of the QWs and thin Mn-containing layers. These investigations show absence of Mn atoms inside the QW. The quality of the structures was also characterized by photo-luminescence spectra from the QWs. Transport properties reveal features inherent to ferromagnetic systems: a specific maximum in the temperature dependence of the resistance and the anomalous Hall effect (AHE) observed in samples with both 'metallic' and activated types of conductivity up to ~ 100 K. AHE is most pronounced in the temperature range where the resistance maximum is observed, and decreases with decreasing temperature. The results are discussed in terms of interaction of 2D-holes and magnetic Mn ions in presence of large-scale potential fluctuations related to random distribution of Mn atoms. The AHE values are compared with calculations taking into account its 'intrinsic' mechanism in ferromagnetic systems.

## 1. Introduction

Opportunity for creation of the spintronic devices has made investigations of diluted magnetic semiconductors (DMS) one of the most vivid directions of current solid-state physics [1-3]. DMS are semiconductors, which contain up to 10% of magnetic impurities. There exist a number of experimental results [2,3] on ferromagnetism and its influence on transport of charge carriers in *p*-type III-V semiconductors doped with Mn. At high concentrations up to 10$^{21}$ cm$^{-3}$ Mn in these compounds has acceptors properties leading to the appearance of local magnetic moments together with free charge carriers [2,3]. Although the microscopic mechanism of magnetic ordering in these materials is still under discussion, it is generally accepted that ferromagnetism would be mediated by free and localized holes in the impurity band. In spite of the fact that there exist a large number of publications dedicated to (III,Mn)V DMS, investigations of two dimensional structures are still relatively rare [4-8].

In the present work the correlation of structural and transport properties of ferromagnetic GaAs/δ-Mn/GaAs/In$_x$Ga$_{1-x}$As/GaAs QW structures with high Mn content are investigated. In the majority of previous investigations of two-dimensional DMS showing the anomalous Hall effect, the Mn layer was located in the two-dimensional electron channel [4-7], or in spite of the spacer and rather low Mn content (0.6 ML [4]), some amount of Mn or defects had penetrated into the channel as indicated by low charge carrier mobility (2-5 cm$^2$/(V·s) [4, 5]). One of the goals of the present article is to investigate the influence of Mn distribution on transport properties of the structures in which, opposite to earlier investigations, the QW and the Mn layer are well separated from each other resulting in high carrier mobility (2000 cm$^2$/(V·s)), while influence of magnetic ordering on the transport properties is still preserved. X-ray and photoluminescence (PL) studies are used to evaluate the Mn distribution and to prove the high quality of the QWs free from Mn atoms. Here we report peculiarities caused by magnetic ordering in transport properties of GaAs/δ-Mn/GaAs/In$_x$Ga$_{1-x}$As/GaAs QW structures. The observed anomalous Hall effect conductivity is in good agreement with the theoretical predictions that the "intrinsic" mechanism [9, 10] is the main reason for AHE in this case. The unexpected non-monotonic temperature dependence of AHE correlates with results of the structure studies of our samples.

## 2. Samples

The samples containing an In$_x$Ga$_{1-x}$As QW of width $d$ = 10 nm inside a GaAs matrix were grown by MOS-hydride epitaxy. The *p*-type conductivity in the well was achieved by δ-doping from the buffer side with a carbon layer, separated from the QW by a spacer of thickness $d_s$ = 10 nm. A Mn δ-layer separated from the QW by a 3 nm thick spacer was prepared by laser deposition [8]. The buffer layer and the spacers were grown at the temperature of 600 °C, while the deposition of the Mn and cap layers was performed at 450 °C. Schematic cross-section of the samples is shown in Fig.1. The parameters are given in Table 1.

As shown by previous investigations [8], the optimal spacer width $d_s \approx 3$ nm and there exists an optimal magnetic impurity concentration for strongest influence of ferromagnetism on the transport properties of the 2D channel. This spacer thickness is connected, most probably, with nonzero thickness of the Mn δ-layer, since according to our previous studies at $d_s < 3$ nm drastic changes in the PL spectra and the activation energy of the conductivity occur, suggesting penetration of the Mn atoms into the QW. Samples with Mn atoms that penetrated or located inside the 2D conductivity channel were investigated in [4, 5] by detecting the magnetic order with AHE. The low values of the Hall mobility of the carriers (less than 10 cm$^2$/(V·s)) support the statement that Mn or defects had entered the 2D channel. Both, the structural investigations and the high value of the Hall mobility ≈ 2000 cm$^2$/(V·s) at 4.2 K give evidence that in our samples the Mn ions are located outside the 2D channel. Increasing of the distance of the Mn δ-layer from the QW leads to weakening of AHE [8], i.e. of magnetic ordering because the ferromagnetism in such systems is based on carrier mediated exchange interaction.

The optimal Mn concentration in the δ-layer can be attributed to the Mn atoms at low concentration substituting for Ga and acting as acceptors, while at high concentrations (above 6 at.% in bulk samples) a considerable part of Mn atoms are located in interstitial sites playing the role of double donors [3,11]. Also at high Mn concentrations formation of MnAs clusters is possible [12, 13], giving a considerable contribution to the magnetization of the whole structure, but not to AHE because of the Shottky barrier between the free charge carriers and the clusters [14]. This conjecture is confirmed by comparison with our previous data of magnetization and Hall measurements on samples with different contents of Mn in the δ-layer [8]. Contrary to [8] in the present structures room temperature magnetization was not observed as it should be at absence of MnAs grains [4, 5 11-13].

Our present samples demonstrate, depending on the growth conditions, both non activation (quasi-metallic, sample A) and activation (sample B) types of conductivity (see section 4.2). For comparison, data for the sample C, which instead of Mn contains a carbon δ-layer, are presented.

Transport measurements were performed at $T$ = 4.2-80 K in fields up to 14 T with a Hall bar sample (width $W$ = 0.3 mm, distance between potential probes $L$ = 1.5 mm) prepared by photolithography, while for X-ray studies samples of size 3x20 mm$^2$ were used.

## 3. High-resolution X-ray diffractometry and reflectometry

X-ray diffraction measurements of the samples were carried out at a double-crystal spectrometer using CuK$_{\alpha 1}$ radiation and quasi-nondispersive scheme for [15, 16]. High quality Ge (400) and Si (111) crystals were used as collimators in the diffractometry and the reflectometry experiments, respectively. The X-ray rocking curves versus $\Delta\theta = \theta - \theta_B$ (where $\Delta\theta$ is the deviation of the incidence angle $\theta$ from the precise Bragg's angle $\theta_B$) were measured using $\theta/2\theta$ scans with a horizontal slit positioned in front of the detector in order to minimize diffuse and background scattering. The obtained diffraction curves are shown in Fig. 2 with vertical dashes indicating the statistical error at each spectral point. For all samples the intensity of the "tails" of the rocking curve is higher at angles smaller than the exact Bragg angle for the GaAs substrate, corresponding to the larger lattice parameter of the In$_x$Ga$_{1-x}$As QW and giving an opportunity to estimate the In concentration ($x$). There are also a number of oscillations in the rocking curves, as typical for high-quality multilayer structures with relatively abrupt interfaces between the layers [15-18]. These interference patterns are mainly due to coherent shift of atomic planes of the relatively thick GaAs cap layer with respect to those of the substrate, induced by the $In_xGa_{1-x}As$ QW. Such a shift results in a phase shift between the amplitudes of the diffracted waves from the cap layer and the substrate [15];

$$\Phi \propto \frac{\Delta a_\perp}{a} l_{QW}, \qquad (1)$$

where $\Delta a_\perp/a$ is the relative lattice mismatch regarding to substrate in the growth direction and $l_{QW}$ is the thickness of the QW.

Following [18] a multilayer structure can be regarded as a system of homogeneous layers with the number of the layers is not necessarily equal to that of the layers grown under the applied technological conditions. Each layer $j$ is characterized by the thickness $l_j$, the lattice mismatch $\Delta a_{j\perp}/a$ and static Debye-Waller factor $w_j$ which defines the degree at amorphism of the layer $f_j$, defined by the reciprocal lattice vector $K$ and chaotic displacement $u$ of the atoms from their regular positions into the $j$-th sublayer

$$f_j = \exp(-w_j) = \exp(-\langle (Ku)^2 \rangle_j), \qquad (2)$$

The structural parameters were found by the conventional least-squares method using the growth parameters of the heterostructures as initial approximation. However, a pronounced inconsistency between the experimental and the calculated rocking curves indicates that the real structures differ from those specified by the growth conditions. In order to improve the quality of fitting of the experimental rocking curves, we had to introduce additional sub layers describing the interfaces on both sides of the QW and the natural distortions on the sample surface. The results of fitting with a seven-layer model are shown in Fig. 3.

The most interesting result of the X-ray measurements is that in sample **A** the GaAs spacer between the QW and the Mn δ-layer is relatively well formed while in sample **B** this spacer prepared with practically same technological parameters is strongly doped with Mn (Fig. 3). This obstacle influences the transport properties of the samples as will be shown below. Existence of a well defined δ-layer of carbon in the sample **C** is questioned because such a light element as carbon is likely to diffuse into the surrounding layers.

The parameters $\Delta a_\perp/a$ and $l$ for such thin layers are strongly correlated so that they could not be simultaneously determined with good accuracy. In particular, the rocking curve for the sample **A** as shown in Fig. 2 can be well fitted both within the model shown in Fig. 3 and a model with 'technological' δ-Mn layer. In order to resolve the ambiguity one should apply additional methods of structural characterization.

Specific feature of our heterostructures is that the electron density of the GaAs matrix differs from that of the $In_xGa_{1-x}As$ QW by only a few tenths of percent while the electron density of the Mn layer is about twice larger. Due to this fact, more detailed information about structural characteristics of thin Mn layers can be obtained by X-ray reflectometry because the amplitude of mirror reflection is mainly defined by depth distribution of the electron density. The glancing-incidence X-ray rocking curves from shown in Fig.4. demonstrate oscillations of the reflection intensity although not so pronounced as those in the diffraction curves in Fig. 2. However, the former are due mainly to coherent shift of the atomic planes of the GaAs cap layer with respect to those of the substrate, which is defined by the thin Mn layer. The amplitude of the oscillations is proportional to the difference between the electron densities in GaAs and Mn [19], the corresponding beats being modulated by a smoothly varying phase:

$$\Phi(\theta) \propto l_{Mn}\theta, \qquad (3)$$

where $l_{Mn}$ is taken for Mn layer thickness. These circumstances reduce the phase problem (1) for very thin layers. Even qualitative analysis of the curves in Fig.4 shows that the reflectometry data can be fitted only by assuming that a thicker layer of Mn atoms 'diluted' in GaAs but not with a 'technological' $\delta$-Mn layer.

We have analyzed the experimental glancing-incidence X-ray rocking curves with a formalism [19] based on Parrat recursion relations for calculating the amplitude of X-ray reflection [20] within continuous depth distribution of the electron density in the interfaces between the layers ($0<z<b_j$),

$$\rho_{j+1}(z) = \rho_j + (\rho_{j+1} - \rho_j)\frac{b_j - z}{b_j}. \qquad (4)$$

The results agree with those evaluated from the diffraction data and allow the total content of Mn in the layers to be estimated as $d_{Mn}$ = 0.27(5) ML and 1.2(4) ML for the samples **A** and **B**, respectively. These values agree qualitatively with the growth conditions (Table 1). At the same time the results give evidence that the Mn atoms did not penetrate into the QWs of the samples **A** and **B**.

**4. Photoluminescence**

The high structural quality and absence of Mn in the QW are confirmed by the photoluminescence (PL) results in Fig. 5, made at 77 K under illumination with a 40 mW He-Ne laser. These spectra show two maxima, the left one is related to the InGaAs QW because with increasing In content it shifts to lower energy and the right one to transitions in GaAs. The transition energies are in agreement with those calculated for the electrons ($E_{e1}$) and heavy holes ($E_{hh1}$) in the GaAs/InGaAs/GaAs QW [21]. Using the *x* values obtained from X-ray results (see Table 1) the calculated transition energy ($E_{e1}$ - $E_{hh1}$) is 1.25 eV for sample A and 1.31 eV for sample B in accordance with the PL results (1.27 and 1.33 eV, respectively). The difference between the observed and calculated values can be related with the fact that [21] does not take into account the stress caused by lattice mismatch. The weakening of the stress by structure distortions in the spacer layer (see Fig. 3) supports this consideration. However, it should be mentioned that the difference in the PL peak positions for structures A and B ($\approx$0.06 eV) coincides with the calculated value within 0.001 eV.

Good agreement between the calculated and measured PL peak energy proves the essential role of size quantization and 2D nature of the charge carriers. Also the Pl signal from QW has a single maximum without low energy satellites originating in DMS from recombination of electrons with holes bound to Mn acceptors [22] in agreement with absence of Mn inside QW. So, both the X-ray and PL results prove the high quality of the QW and confirm the absence of Mn atoms inside it. Also a substantial difference between the Mn distribution and spacer the homogeneity is demonstrated in samples with high (**B**) and low (**A**) Mn content, resulting in sharp distinctions of their transport properties as it will be seen below.

## 5. Transport properties

The hole mobility $\mu_p \approx 2000$ cm$^2$/Vs in our samples exceeds more than two orders of magnitude those in the 2D A$^{III}$B$^V$ structures doped with Mn (< 10 cm$^2$/Vs), and where AHE was observed [4,5]. The high values of $\mu_p$ in our samples may be connected with good interface and absence of the Mn atoms in the conductivity channel as shown by the X-ray and by PL results. Random distribution of Mn could result in carrier localization in fluctuation potential (FP) wells, that may be one of the reasons for the similarity of $p_s$ in the **A** and **B** samples in spite of the different doping levels $d_{Mn}$ (see Table 1). The FP arising at high Mn contents will be discussed below.

As shown in Fig. 6. in the sample **B** with high Mn content (high amplitude of FP) the resistance $R_{xx}$ grows exponentially upon cooling, while in the sample **A** with low Mn content the resistance changes with temperature less than 30% (curve A in Fig. 6). Temperature dependence of $R_{xx}$ in **A** resistance is nonexponential (activation energy is zero). This dependence is similar (except of the hump) to that in analogous structures without the Mn δ-layer and well described by quantum corrections to conductivity [23]. This statement is supported by analysis of the magnetoresistance and the temperature dependence of the conductivity. For simplicity we will refer sample **A** as having quasi-metallic conductivity.

The random distribution of the charged Mn ions both in the sample plane and along the thickness of the structures can give rise to FP. One of the main reasons for FP is the compensation accompanying with high level of doping [24], another is the fluctuation of the Mn layer thickness that results in fluctuations of the effective spacing between the Mn layer and the QW. In the frame of the FP model the unexpected transition from quasi-metallic (**A**) to activation temperature dependence of conductivity (**B**) with increasing doping could be naturally explained. The reason is that the fluctuation potential will show itself in sample **B** with higher Mn content more distinctly than in sample **A,** leading to the carrier localization. It is clear from the X-ray measurements (Fig. 3) that fluctuation of the Mn layer thickness (penetrating of Mn atoms inside the spacer) is much stronger in sample **B** than in **A**.

The compensation level in sample **B** is also higher than in **A** due to the reason that above certain concentration (6 at % in bulk, not annealed samples [11-13]) the Mn atoms enter interstitial states rather than Ga positions [3]. Because the Mn ions substituting for Ga act as acceptors and those occupying the interstitial sites as double donors [3, 11, 12]. The total concentration of the charged impurities, $N = N_A^+ + N_D^-$, strongly exceeds that of the carriers, $p = N_A^+ - N_D^-$, which can not screen random spatial fluctuations of charge $\sim N^{1/2}$. Here $N_D$ and $N_A$ are the concentrations of the donors and the acceptors, correspondingly. The unscreened charge gives rise to long-range FP. So both reasons for FP tend to localize the carriers in the FP potential wells, resulting in activation character of the conductivity in **B**.

Theory of the FP related to compensation in semiconductors is well developed for 2D nonmagnetic structures [25], following the known results for 3D systems [24]. In particular, the case of high impuritiy concentration of ($N \gg a_B^{-2}$) close to the conductivity channel when $d_s < a_B$ was studied, where $d_s$ is thickness of the spacer between the conductivity channel and the plane of the doping layer and $a_B$ is the Bohr radius. This is just the case of our samples where $a_B \approx 6$ nm, $d_s \leq 3$ nm. The distance between the percolation level and the Fermy energy is about the mean square amplitude of the FP wells [25]

$$\gamma = \beta \frac{e^2}{\kappa} N^{1/2} \ln^{1/2} \frac{N^{3/4}}{p a_B^{1/2}} = \beta \frac{e^2}{\kappa} R_c p \ln^{1/2} \left( \frac{R_c^3 p}{a_B} \right)^{1/2}, \qquad (5)$$

where $\beta$ is the factor of about 1, $p$ is the surface density in the conducting channel (both localized and delocalized), $\kappa$ is the permittivity and $R_c = N^{1/2}/p$ is the screening length which characterizes the FP scale [25]. The concentration $N$ of the charged impurities is not well known and usually is much lower than the concentration of introduced Mn impurities [26, 27]. Therefore, it is more instructive to use the relation $R_c = N^{1/2}/p$ and express $\gamma$ in terms of $R_c$ and $p$ – parameters which could be found from the experimental data (see below).

From Eq. 5 it follows, that $\gamma$ does not depend on $d_s$ [25]. This is due to the dependence of the Coulomb energy not only on the distance between the Mn δ-layer and the QW but also on the characteristic size of the carrier wave function and the $d_s$ value is not important if $d_s < a_B, d$. Here $d$ is the QW thickness. So, for the first glance the fluctuations of the Mn layer thickness, which could be treated as fluctuations of the effective spacer thickness, do not affect the energy of the carriers in the FP wells. However in our case $d_s$ is less but comparable with $a_B$ and $d$ and one should take into account the penetration of Mn atoms inside the spacer (see Fig. 3) as an additional reason for FP. This statement is reinforced by existence of the structural defects and interface roughness caused by Mn in the spacer (see Fig.3). However, having no other theoretical results we will use Eq. (5) for rough evaluation of the FP parameters.

The charged carriers are localized at the minima of the fluctuation potential where they have high densities forming "metallic" droplets. We call them metallic because of the long-range nature of FP since when the size of the potential well exceeds the mean free path of the holes they can act as free carriers. At the same time the Mn concentration in these regions is above average. Both high density of the carriers and the Mn impurities promote magnetic ordering and possible local ferromagnetic transition in these areas [28] since the ferromagnetic ordering is expected to be mediated by carriers [3]. As the result, the sample contains conductive regions ("metallic droplets or lakes"), which could be ferromagnetic at low enough temperature.

The metallic droplets may form a continuous percolation cluster resulting in quasi-metallic conductivity (sample **A**) or localization of the carriers in the FP potential wells leading to activation type conductivity. The activation character of the sample **B** conductivity is shown in the insert to Fig. 7. At $T > T_t = 24$ K the $R_{xx}(T)$ conductivity is determined by carrier activation over the percolation level with the activation energy $\varepsilon_a = 10.5$ meV and at $T = T_t$ there is crossover to 2D hopping conductivity with $\ln R_{xx} \propto (1/T)^{1/3}$. This is in accordance with theoretical results for transport in FP [25].

The crossover temperature from activated conductivity of the carriers to tunneling between the FP wells can be used for estimation of the characteristic size of the potential well and the radius $R_c$ of the metallic droplets [25]. Following [25] the energy gap between the percolation level and the Fermi level is roughly equal to the amplitude of the fluctuating potential, $\gamma \approx \varepsilon_a = 10.5$ meV. For estimation of $R_c$ we compare the probability of activation of the holes to the percolation level, $w_a \sim \exp(-\gamma/kT)$, with the probability of tunneling between the hole droplets, $w_t \sim \exp(-2R_c/\lambda)$, where $\lambda = \hbar/(2m^*\gamma)^{1/2}$ is the decay length of the wave function under the barrier of height defined by the energy difference between the percolation and the Fermi levels [25]. Taking into account that crossover takes place at $T = T_t = 24$ K one gets $R_c \approx 20$ nm, which is longer than the mean free path of the holes $l_h$. At $\mu \approx 10^3$ cm$^2$/V·s and $T \approx 30$ K we get $l_h \approx 6$ nm. Using the obtained value $R_c \approx 20$ nm and the holes concentration $p \sim 10^{12}$ cm$^{-2}$ on the base of Eq. (5) one can estimate $\gamma \sim 10 - 20$ meV, that is in agreement with $\varepsilon_a = 10.5$ meV.

The main peculiarity of the data in Fig. 6 is the maximum of $R_{xx}(T)$ observed in the samples **A** and **B** at about 30 K, but not in the carbon containing sample **C**. Existence of a maximum in $R_{xx}(T)$ is common for 3D and 2D DMS structures and ferromagnetic metals as well [4, 12]. This feature is often used for estimation of the Curie temperature [3, 4, 12], also in the case of an activation type conductivity [29]. Accordingly this gives in our samples the value of $T_c \approx 30 - 40$ K. Exact description of the resistivity hump in DMS is under discussion and several models have been suggested, taking into account scattering of carriers by critical fluctuations close to the ferromagnetic transition, splitting of the hole spin subbands and so on [3, 12]. In highly disordered materials the peculiarities of $R_{xx}(T)$ at $T = T_c$ could be caused by phase separation and carrier scattering by magnetic clusters [30, 31]. Analogous effects could take place in our case when the holes are scattered by magnetic moments of the metallic droplets and/or due to increasing of the size of the droplets with lowering the temperature [28].

More detailed information on magnetic properties in particular, on $T_c$, can be obtain by measurements of AHE [3, 4, 12]. The fluctuating model featured above helps us to understand details of AHE presented in Fig. 8. Non-linearity of the Hall resistance is typical for ferromagnetic materials where the Hall voltage is a sum of normal and anomalous contributions and the Hall resistance

$$R_{xy} = \frac{\rho_{xy}}{d} = R_{xy}^n + R_{xy}^a = \frac{R_0}{d}B + \frac{R_s}{d}M, \qquad (6)$$

where $d$ is the thickness of the conducting channel (QW in our case), $R_{xy}^n$ and $R_{xy}^a$ are normal and anomalous contributions to the Hall resistance, while $R_0$ is the constant of the normal Hall effect, $R_s$ is the AHE constant caused by spin-orbital interaction of carriers and related to exchange splitting of the spin sub bands of the holes proportional to the magnetization $M$. Hence AHE arises from interaction between charge carriers and the magnetic subsystem and could be significant up to (2-3) $T_c$. [11, 12].

In 2D structures measurements of AHE is an effective method for studying magnetic ordering, because small values of $M$ cannot be extracted by magnetometer measurements due to large diamagnetic contribution from the substrate [4, 7]. In our case, the normal component $R_{xy}^n$ in Eq. 6 strongly predominates and special efforts are needed to extract the AHE component. We determined $R_{xy}^n$ in a high magnetic field taking into account that contrary to the AHE this component is not saturated in magnetic field. The results are presented in Fig. 8. The data for the sample **A** is given at $T = 17$ K, because at higher temperatures the ratio $R_{xy}^a / R_{xy}^n = \rho_{xy}^a / \rho_{xy}^n$ is too small. It should be emphasized, that this is the first observation of AHE in 2D DMS structures with high carrier mobility. The small value of $\rho_{xy}^a / \rho_{xy}^n$, is not due to small AHE conductivity, $\sigma_{xy}^a$, but is owing to the high value of the normal component of the Hall effect, $\sigma_{xy}^n$.

The values of the anomalous and the normal Hall conductivities are

$$\sigma_{xy}^a = \rho_{xy}^a / \rho_{xx}^2 \text{ and } \sigma_{xy}^n = \rho_{xy}^n / \rho_{xx}^2, \qquad (7)$$

In a not too strong magnetic field we have $\sigma_{xy}^n = \frac{pe^2}{m^*}\omega_c\left(\frac{\tau^2}{1+\omega_c^2\tau^2}\right) \propto \tau^2$ assuming that $\omega_c\tau < 1$. Here $m^*$ is the hole effective mass, $\tau$ is the momentum relaxation time and $\omega_c = \frac{eB}{m^*c}$ is the cyclotron frequency. On the other hand, according to recent theoretical results the main mechanism contributing to AHE in DMS and, particularly, in magnetic 2D structures [9, 10] is "intrinsic" or non-dissipative [3, 32, 33], meaning that $\sigma_{xy}^a$ does not depend on scattering rate. So as it follows from (7) the ratio $\rho_{xy}^a / \rho_{xy}^n = \sigma_{xy}^a / \sigma_{xy}^n \propto \tau^{-2}$ and the contribution of AHE to the signal is higher at lower carrier mobility. That is why AHE in 2D DMS structures was observed up to now only in structures with the very low carrier mobility. This is the reason why $R_{xy}^a$ for the sample **B** is considerably smaller than measured by Nazmul et. al. [4] in similar structures with mobility about two orders of magnitude less. At low temperatures the carrier mobility in samples **A** is several times greater than in **B** and $R_{xy}^a$ is correspondingly smaller (see Fig. 8), while $\sigma_{xy}^a$ is of the same order of magnitude.

Finally, we get $\sigma_{xy}^a \cong 0.07 e^2/h$ for sample **A** and $\sigma_{xy}^a \cong 0.17 e^2/h$ for sample **B** in agreement with recent theoretical calculations of AHE in 2D structures [9, 10]. The difference between these values correlates with different Mn contents (see Table 1) and the depth of Mn penetration into the spacer in the **A** and **B** (see Fig. 3). From the data in [9, 10] one can obtain $\sigma_{xy}^a \sim 0.1 e^2/h$ in accordance with the measured values. So the experimental results support the theoretical prediction that the "intrinsic" mechanism is the main reason for AHE in our samples. Additional support for this could be obtained by comparison of the theoretical and experimental results for the dependence $R_{xy}^a(\rho_{xx})$. For the "intrinsic" mechanism the relationship $R_{xy}^a \propto \rho_{xx}^2$ is just what we have found for sample **B** using the temperature as the parameter (insert in Fig 9). Of course, this result is valid only at high enough temperatures, when the conductivity is determined by the carriers activated above the percolation level but not by the hopping mechanism.

It is widely accepted that observation of AHE is the main tool for establishing magnetic ordering in DMSs [3], especially in 2D structures for where bulk magnetization measurements are difficult. The present results provide evidence of magnetic ordering and hole mediated coupling between local Mn moments in our samples, in accordance with theoretical calculations [9, 10] of AHE in 2D DMS structures assuming substantial spin polarization of the carriers. So we can conclude that very likely the carriers in our samples are spin polarized, as also supported by observation of photoluminescence polarization in similar structures [34].

The temperature dependence of $R_{xy}^a$ suggests that AHE can be observed for sample **B** up to $T_m^*$ ≥ 80 K, which and, correspondingly, the spin polarization of the carriers agrees with the value $T_m \approx$ 70÷100 K below which the resistivity starts to grow (see Fig. 6). Because the AHE and the magnetic moment should be measurable below $T \approx (2-3)T_c$. [11, 12] we get $T_c \approx 30 - 40$ K in agreement with the temperature of the maximum in $R_{xx}(T)$. According to the FP model, $T_c$ is likely related to local ferromagnetic transitions inside the metallic hole droplets. Below 40 K the value of $R_{xy}^a$ for sample **B** surprisingly drops down. The reason for this is the crossover to hopping conductivity with lowering of the temperature. It should be noted that at first approximation the hopping conductivity does not contribute to the Hall effect. So AHE drops with diminishing of the activated transport as can be observed from Fig. 9. In this case the Hall voltage arises only inside droplets with high concentration of holes.

As shown in Fig. 8 the magnetic field dependence of $R_{xy}^a$ for sample **A** gives some hints for existence of a hysteresis loop. However, the small signal to the noise ratio prevents from making any strong conclusions. On the other hand, the hysteresis loop should be weakened because the anisotropy of the sample tends to align the magnetic moment along the plane of the structure, diminishing the Hall effect around zero magnetic field.

**6. Conclusions**

As a result of versatile investigations, we have observed strong correlation between structural, electronic and optical properties in 2D GaAs/Ga$_{1-x}$In$_x$As/GaAs QWs containing a Mn δ-layer. With X-ray difractometry and reflectometry measurements precise distributions of the lattice parameters of the QWs and the Mn δ-layers are obtained. The structural investigations confirm absence of Mn atoms inside the QW. The quality of the sample structures is characterized also by photo-luminescence spectra showing emission lines from the QWs in good agreement with X-ray mesurements.

Although there should be no Mn atoms inside the QW the presence of AHE and the maximum in the temperature dependence of the resistivity component $R_{xx}$ at 35 K give evidence for magnetic ordering in samples **A** and **B**. As a result, the AHE is observed for the first time in a 2D DMS structure with high mobility of the holes (2000 cm$^2$/Vs). The values of $\sigma_{xy}^a$ agree reasonably with theoretical predictions [9, 10] in corresponding structures in presence of spin polarized carriers. The agreement between the measured and calculated values of $\sigma_{xy}^a$ and the observed $R_{xy}^a \propto \rho_{xx}^2$ dependence are ymptoms of "intrinsic" AHE in our samples. The $\sigma_{xy}^a$ values also correlate with the Mn content of the samples.

The conductivity at a high Mn content (1.5 ML) is determined by hopping of carriers localized in the wells of long-range fluctuating Coulomb potential. The amplitude and the scale of the fluctuations are estimated from temperature dependence of the resistivity with the result that the size of the potential wells exceeds notably the mean free path of the holes. With lowering of the temperature AHE reaches a maximum at $T = T_c$ and diminishes with further decrease of the temperature.

The authors are grateful to K.I.Kugel and E.Z. Meilikhov for the fruitful discussions, to ISTC (grant G1335), RFBR (grants 05-02-17021, 07-02-00927) and the Wihuri foundation for support.

Table 1. Technological and physical parameters of the samples (M stands for metallic and I for activation type of conductivity).

| Sample | $x$ (X-ray) | $d_{Mn}$, ML | 77 K | | |
|---|---|---|---|---|---|
| | | | $\mu_{eff}$, cm$^2$/(V·s) | $p_s \cdot 10^{-12}$, cm$^{-2}$ | $R_s$, Ω/□ |
| **A** (M) | 0.21 | 0.5 | 1860 | 2 | 1660 |
| **B** (I) | 0.16 | 1.8 | 1350 | 1.8 | 2540 |
| **C** (I) | 0.18 | 0 | 1598 | 0.5 | 7800 |

Figures Captions

Fig. 1. Configuration (a) and the band diagram (b) of the investigated samples.

Fig. 2. X-ray rocking curves taken from (004) planes of GaAs/$\delta$-Mn/In$_x$Ga$_{1-x}$As/GaAs (vertical dashes) and corresponding theoretical curves (solid lines) calculated within a seven-layer model (the mean square standard deviation $\chi^2$ = 1.45, 1.04 and 1.15 for *A*, *B* and *C* samples, correspondingly). Noncoherent diffuse background is shown by dotted lines.

Fig. 3. Depth (*z*) profiles of the lattice mismatch evaluated from fitting the experimental X-ray rocking curves.

Fig. 4. Experimental reflectometry curves from GaAs/$\delta$-Mn(C)/In$_x$Ga$_{1-x}$As/GaAs heterostructures (vertical dashes) and theoretical curves (solid lines) calculated (the mean square standard deviation $\chi^2$ = 1.39, 5.22 and 4.68 for *A*, *B* and *C* samples, correspondingly).

Fig. 5. Photoluminescent spectra of the GaAs/$\delta$-Mn(C)/In$_x$Ga$_{1-x}$As/GaAs samples *A*, *B* and *C*. The maximum on the right hand side originates from GaAs and that on the left hand side from the QW.

Fig. 6. Temperature dependence of the resistivity for the GaAs/$\delta$-Mn(C)/In$_x$Ga$_{1-x}$As/GaAs samples **A, B** and **C**.

Fig. 7. Temperature dependence of the resistance of sample **B**. The insert shows the resistance vs. 1/*T*. At low temperatures $1/T^{1/3}$ provides better fitting than 1/*T*, while at higher temperatures the situation is quite the opposite.

Fig. 8. Magnetic field dependencies of $R_{xy}^n$ (upper panels) and of $R_{xy}^a$ (lower panels) for the samples **A** (T = 17 K) and **B** (T = 55 K). The value of the anomalous component was obtained by deduction of the normal Hall component estimated in fields up to 15 T from the total Hall response.

Fig. 9. The temperature dependence of the anomalous Hall resistance $R_{xy}^a$ for sample **B**. The insert demonstrates the parametric dependence between $R_{xy}^a$ and $R_{xx}$.

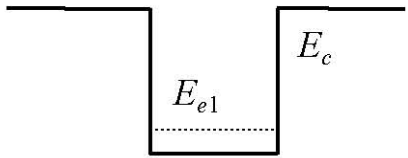

(a)

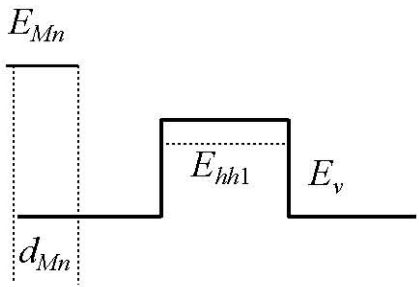

(b)

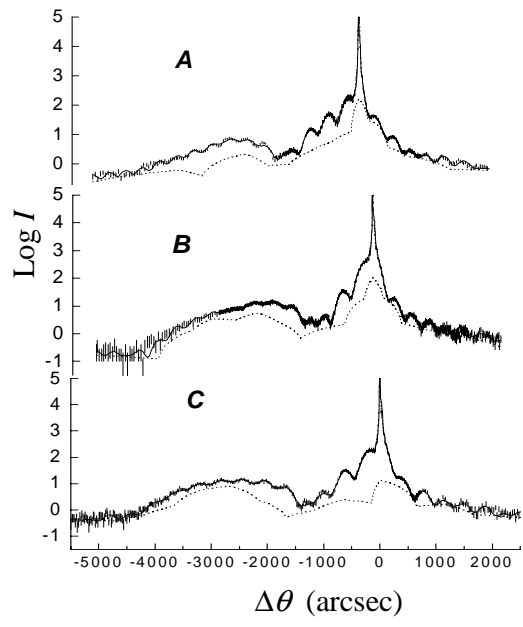

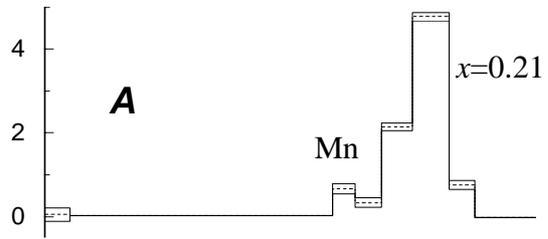
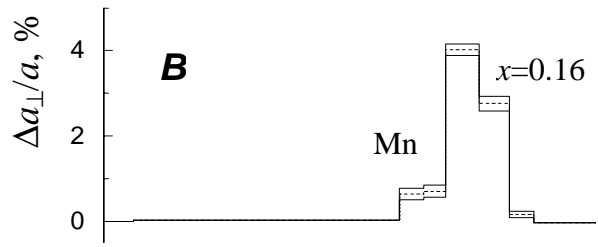
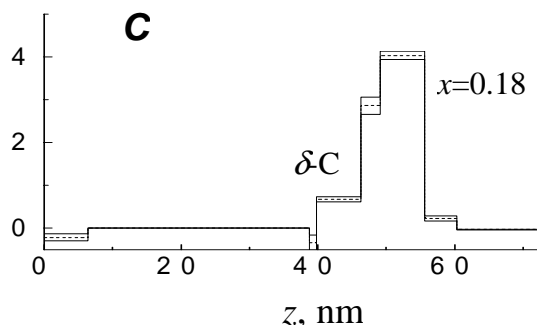

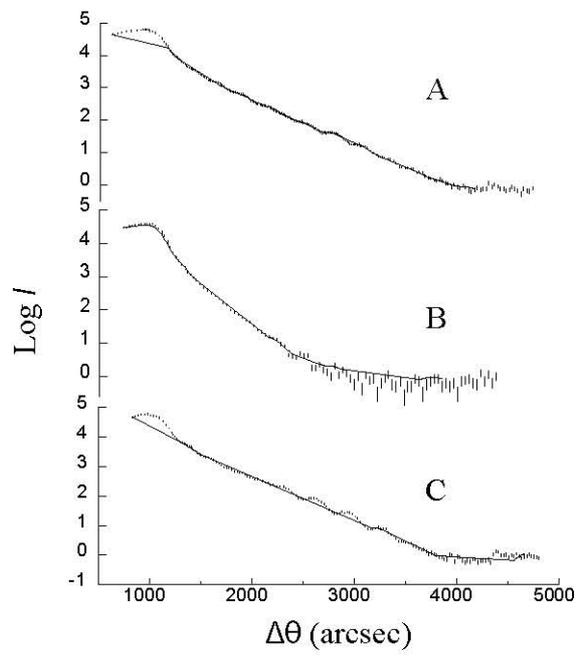

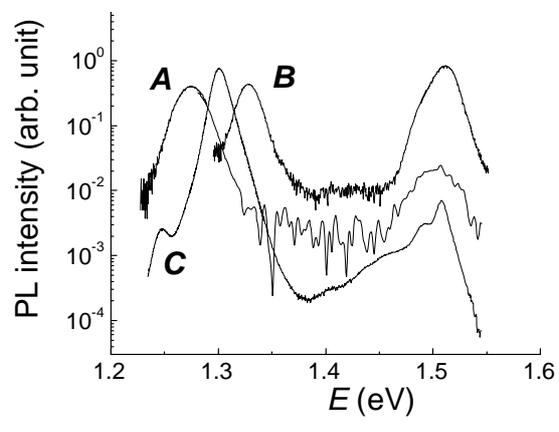

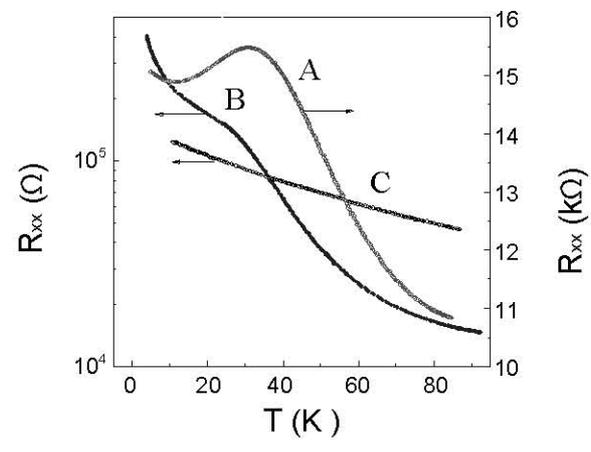

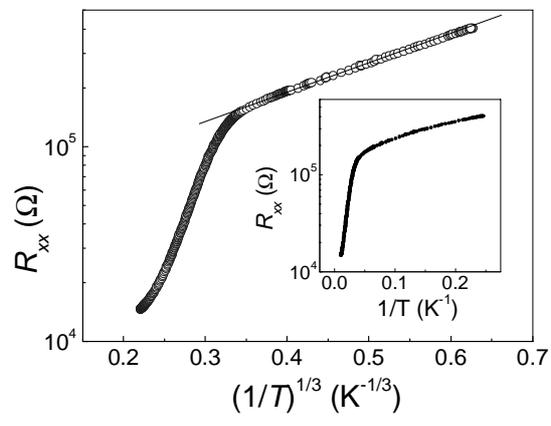

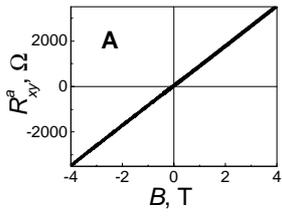 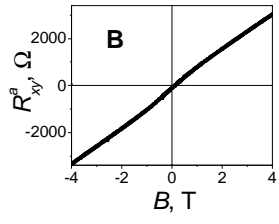
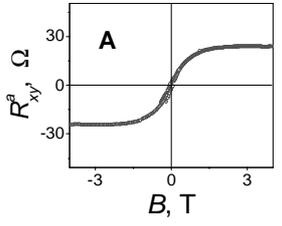 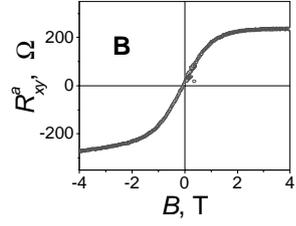

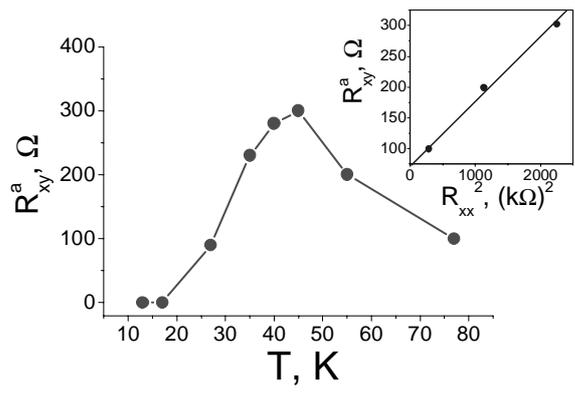